\documentclass[12pt,preprint]{aastex}
\begin{document}
\title {BLACK HOLE OR MECO? DECIDED BY A THIN LUMINOUS RING STRUCTURE
DEEP WITHIN QUASAR Q0957+561}
\author{Rudolph E. Schild\footnote{Center for Astrophysics, 60 Garden
Street,
Cambridge, MA 02138, USA: rschild@cfa.harvard.edu; corresponding author,
phone 617-495-7426: FAX 617-495-7467} \& Darryl J. 
Leiter\footnote{Marwood Astrophysics 
Research Center, Charlottesville, Virginia 22901, USA: dleiter@aol.com}}
\begin{abstract}
Optical, Infrared, X-ray, and radio wavelength studies of quasars
are beginning to define the luminous quasar structure from techniques of
reverberation and microlensing. An important result is that the inner
quasar structure of the first identified gravitational lens, Q0957+561 A,B
seems not to show the kind of structure expected for a supermassive black
hole, but instead show a clean-swept interior region as due to the action
of a magnetic propeller, just as expected for a MECO (Magnetic Eternally
Collapsing Object) structure. Given the present state of the observations,
the strongest model discriminant seems to be the existence of a thin luminous
band around the inner edge of the accretion disc, at a distant radius $\sim$
$70 R_G$ from the $\sim$ $4\times 10^9 M_\odot$ central object. 
Since the existence of a clean magnetic propeller swept inner region 
$\sim$ 70 $R_G$ surrounded by a sharp $\sim$ 1 $R_G$ disc edge
are the low-hard state spectral properties
associated with a highly redshifted central MECO object, we are led to the
conclusion that these observations imply that the Q0957 quasar contains a
central supermassive MECO instead of a black hole. In this report we review
the details of the observations which have compelled us to reach 
this conclusion.
\end{abstract}
\section{Introduction}

Studies of the inner parsec region of quasars by optical reverberation
and microlensing have shown an important discrepancy when the observations
are compared to a standard Black Hole model. Instead of the expected inner
accretion disc edge at $6R_G$ where $R_G$ is the gravitational radius,
the data seem to evidence
structure indicating a cleanly swept inner region out to a central distance
$\sim 70 R_G$, surrounded by a sharp luminous band of radial thickness 
$\sim 1 R_G$ at the inner edge of the accretion
disc, more typical of the low-hard state of a Magnetospheric Eternally
Collapsing Object (MECO), Robertson and Leiter (2002, 2003, 2004, 2005)
than that of a black hole.
In the report
detailing this conclusion, Schild, Leiter, and Robertson (2006; SLR06) 
presented in their Fig. 2 a schematic diagram which traced out the physical
processes which contributed to this result.

To fortify this conclusion based on these results, we present in this
report a summarized compendium of the results of published investigations
in order to make a comparison of relevant observations to competing
theoretical models. This is necessary because many of the
observational results described in SLR06
are scattered over a large literature focused on
other topics, such as the nature of the microlensing signature of the
Baryonic dark matter, and the time delay implications for determination of
the Hubble Constant. Recall that the Q0957+561 A,B quasar lens system, the
first discovered multiple image gravitational lens system (Walsh, Carswell, 
and Weymann, 1979), was heavily observed in the two decades since its
discovery with the expectation that measurement of its time delay would
allow determination of the Hubble constant by the Refsdal (1964) method,
independent of the classical methods based upon Cepheids. With the discovery
of the time delay by Schild and Cholfin (1986) an unexpected rapid
microlensing was found (Schild 1996), and since the baryonic dark matter
implications eclipsed the time delay and Hubble constant program, a long
literature that followed (Schild 2004a, 2004b, 2007, Schild \& Dekker, 2006)
emphasized the rapid microlensing.

But the same measured brightness record had implications for the nature of
the luminous quasar structure as well. Already the first Q0957 rapid 
microlensing report by Schild (1996) concluded that the quasar must be
larger than previously believed, and "dominated by rings or shells." A
double-ring model simulation
by Schild and Vakulik (2003) showed that the amplitudes
and the durations of the observed microlensing brightness fluctuations on
long and short time scales could be understood to be compatible with
reverberation and
autocorrelation measurements. 

The structure of this paper is as follows. In sections 1-5 we discuss 
the basic
statistics defining the accuracy of the data, followed by a detailed
discussion about
the reverberation data used to infer the location of the Elvis outer
outflow wind structure, and the signature of a sharp ring at the inner edge
of the accretion disc. Then in sections 6 and 7 we conclude by showing
that, instead of a black hole driven structure, the observational results
for Q0957 favor structure dominated by a central dipole field of a MECO
object in a low-hard state. 
-
\section{The Q0957 Brightness Monitoring Data Quality and Quantity}

We begin with a discussion about the quality of brightness monitoring
data. The analyzed data table, available online at
http://www.cfa.harvard.edu/~rschild/fulldata2.txt reduces all data taken on 
an individual calendar night and averages the
data together, and assigns an observation time from the mean for the
individual observations. Data are almost always averages of 4 CCD image
frames analyzed for brightness separately. This usually gives 4 nightly 
brightness
estimates, which usually have a 1-sigma rms near .01 mag, or 1\% brightness
error for the nightly mean. Independent confirmation of this estimate was
made by statistician David J. Thomson, who estimates nightly average 
errors of 0.006 and 0.008 mag 
for the A and B images, respectively (Schild and Thomson, 1995). A
comprehensive analysis of the statistics of the data set focused on the
time delay controversy is given in Thomson and Schild (1997). 

Figure 1 is a plot showing the number of nights where
there is data overlap between images A and B after correction for
cosmological  time
delay. This includes interpolations of up to 2 days, meaning suppose I make
an observation on Jan 1 2000. Then do I have an observation for Jan 1 + 417
days in 2001, with the possibility that on Jan 1 + 417 days I actually made
the observation on that date, or I could interpolate to that date with at
most a 2-day interpolation window? This makes sense, where our sampling rate
is 1 day so the Nyquist frequency is 2 days. We can still expect that the
data point compared to image A on Jan 1 2000 will have a magnitude estimate
accurate at the 1\% level for 417 days later, after the Q0957 structure 
function given by Colley and Schild (2000).

Figure 1 shows how many dates for image A observation will have a
1 \% accurate image B brightness estimate. When this plot was made in 
Jan 1993,
for any date of observation of image A there would be several hundred 1\%
quality measurements for image B made 417 days later. 
We are interested in lags
in this plot of less than 10 nights. The Fig. 1 plot shows that by
1993, at least 270 data points define brightness for lags up to 10 days
from the cosmological 417 days lag.

This means that when I compare the measured brightness of image A with
brightness of image B 417 +/- 10 days later, that brightness is estimated
with high accuracy. Over 200 points overlap. 
Therefore we may conclude that we know very well the brightness properties 
of the two images for
lags near the cosmological lag, 417.1 days (Colley et al, 2003).

The quality of the data, usually expressed as a 1 $\sigma$ error, has been
commented upon repeatedly. Thus Schild and Thomson (1995) reported a median
error of 0.006 magnitudes for both images, and
Schild and Thomson (1997) reported mean error
estimates of (.00954, .01202) for images (A,B). And when Colley and Schild
(2000) compared their refined magnitude estimates with the long Schild et al
record, they found mean rms differences of only (.006, .008) magnitudes
for images (A,B). Because the quasar shows nightly brightness fluctuations
of approximately 1\%, and larger trends on longer time scales, these trends
are well measured with available photometry. The observed nightly brightness
fluctuations for Q0957 have been expressed as a structure function by
Colley and Schild (2000) and as a wavelet amplitude by Schild (1999).
Hence on the basis of the above analysis it is safe to conclude 
that both the quality and the quantity of quasar
brightness data are sufficient to allow meaningful analysis of the
quasar's internal structure to be made.

\section{Outer Quasar Structure Seen in Autocorrelation and Simple Inspection}

The Figure 2 auto-covariance (auto-correlation) plot is expected to show
correlation on time scales related to the outer Elvis structures, around
100 proper days, and structure related to the inner edge of the accretion
disc. In Fig. 2, which was originally shown at the 1993 Padua Symposium
(Thomson and Schild, 1997)
the auto-correlation for the A (dashed) and B (solid line) 
images are shown together. Multiple peaks for time lags near 100 proper 
days are presumed to result from a
disturbance taking place at the inner region of the quasar, near the inner
edge of the accretion disc, and then with reflections or fluoresces off the 
Elvis outflow structures which are expected near to but just within 
the light cylinder radius (Schild and Vakulik, 2003; 
SLR06, SLR07). Because microlensing 
is expected to cause complications in the detection of these outer 
reflections, the B image with its higher microlensing optical depth should 
show different autocorrelation than A. Thus we have included a schematic
"bottom" or "continuum" level in figure 2 for the two quasar images.
This emphasizes how similar the peaks measured for the two lensed images
are in amplitude and width.  Since microlensing can locally amplify 
random parts of the quasar
structure, we consider as significant only peaks found in the
autocorrelation estimates of both images.

The images arising in the lowest-lag Elvis structure, at 129 and 190
days (observer's clock), have about the same width, about 50 days, 
and half again larger widths of 75 days, for 540 and 620 day lag peaks.
The 620 day lag peak is not convincing, by itself, and call it 
unobserved if you
like, but if somebody insists that it must be there, one could argue that
it is. 

A similar pattern of quasar optical reverberations are now recognized 
by direct inspection of the measured brightness curves in a second
lensed quasar system, the Einstein Cross (Q2237), with excellent quality
data (Wozniak et al 2000) available at:

	http://www.astrouw.edu.pl/~ogle/ogle3/huchra.html

In examining the above website the reader is encouraged to scroll down to
see the plot of optical brightness monitoring for the 4
lensed images. The pattern of brightness peaks to be seen are similar in
amplitude and duration to similar peaks found in Q0957. The Einstein Cross
peaks I refer to are best seen in the upper green plot for image A, and the
peaks occurred at (HJD - 24400000) = 2500, 2950, and 3300, with possibly
more, but microlensing makes interpretation insecure. The peaks have a
brightness amplitude of approximately 0.2 magnitudes and durations of
approximately 50 days. The peaks have been discussed in the context of
quasar structure by SLR07. These peaks were used by Vakulik et al (2006) to
determine the time delays of the quasar images.

Comparison of the Q2237 direct brightness curves shows that reverberation
of a central quasar disturbance reflects off the outer quasar structure 
at lags and
time durations about as inferred from the Q0957 autocorrelation estimates.
We find that for both quasars, outer structure gives reverberation
estimates of size at observed scales near 200 days and observed widths near
75 days; with (1+z) cosmological time dilation correction, these become proper
(intrinsic) reverberation delays near 100 days and widths near 25 days.  

Of more interest is the question of whether structure is evident on scales
comparable to the expected structure at the inner edge of the accretion
disk, at $6R_G$. 
This size scale would correspond to elapsed times of less than a day on the
observer's clock. Moreover, in typical autocorrelation estimates,
any significant noise contribution should decrease calculated correlation
away from 0 lag; detection noise and cosmic dispersion
cause longer lags to give strongly decreasing correlation. But here, 
the curves are quite
different; the Q0957 inner structure has the character of wide peaks of
approximately
10 days, with additional sub-peaks at 15 days (A image) and 5 days (B image).
This means that for these short lags the data are correlated, and insofar
as the quasar is bright today it will probably also be bright for 10 days
or so. In other words, an inner structure is in evidence, and insofar as
the autocorrelation peak reflects internal structure limited by causality
at light propagation speed, a structure size scale is implied. Since an
inner scale of 1 light day for the inner accretion disc edge is expected
but not observed, and since instead a broad inner plateau is found, we conclude
that the inner Q0957 quasar structure is not at the location of the innermost
stable orbit (or that quasar brightening effects propagate much slower than
light speed). A further discussion of the nature of this reverberation plot
allowed Schild (2005) to infer the quasar orientation with respect to the
plane of the sky. Even more importantly, SLR06 showed how 
this data implied the existence of a bright narrow band at the inner edge
of the accretion disc, which was required to explain the extremely rapid 
5-hour brightness fluctuations that were seen in the microlensing data.

We present in Fig. 3 a more recent cross-correlation calculation with 
finer output scaling that
shows better the results for 10-day lags, which relate more to the central
quasar structure. This calculation from 1995, with half again as much data,
is for A image alone. The peaks from the outer Elvis structures at 129 days
and 190 days are strong. Their widths are comparable, about 45 days, which
is twice the width of the peak around 0 lag, suggesting that the widths are
dominated by the light scattering or fluorescence from the Elvis
structures, which have larger dimensions, as estimated in SLR06.

It is now extremely significant that the auto-correlation has strong
structure between 0 and 25 day lags. Recall that for noise-dominated data
the data point for 0 lag
should be highest and the autocorrelation function should rapidly decline
for lags of 1, 2, 3, ... 10 days. But the plot shows sub-peaks 
around 6, 11, and 20 days, within a broad
general peak of about 25 days width. This probably is the result of the Q0957
quasar's innermost structure.

Figure 4 shows similar data for the B image. It gives similar results
for the 129 and 190 day peaks, understood to be related to the outer Elvis
structures. The heights and widths are similar for the Elvis peak lags at
129 and 190 days seen in image A (Fig. 3).

And the inner peak again has a width of about 25 days, with again
substructure peaks at 5 and 20 days. The fact that these inner sub-peaks do
not exactly agree probably has to do with microlensing of this inner
structure, which originates in the random star field in the lens galaxy. 

Combining results from Figures 3 and 4 showing autocorrelation estimates 
from 1995, we have the
suggestion of structure on time scales of 6, 11, and 20 days and an overall
size limit for the inner quasar structure of approximately 25 light days
(observed).

\section{Results from Cross-correlation analysis}

The above two plots have been about autocorrelation. Now we wish to discuss
the previously published plots for CROSS-correlation.

First you must ask yourself, with many hundreds of data points of 1\%
quality and with observed nightly brightness fluctuations at least 1\% between 
observations, isn't it
obvious that for some cosmological time delay, say for example, 417 days, 
there will be a
very large spike in the cross-correlation curve? With excellent data
sampling, characterised by hundreds of data points at resolution of 1 day
and with detection noise at or below the amplitude of quasar brightness 
fluctuations, this large spike should have a width of 1 or 2 days.

But this cross-correlation spike has never been found by any of the 10 
research groups that have sought it.

Thus we immediately are confronted with the fact that probably the inner
quasar structure is softening the expected cross-correlation peak.
Fig. 5 is a cross-correlation plot to illustrate this. It was already
published as Fig. 2 in Schild and Thomson (1997). Instead of a large 
spike with a 1-day
half-width, we find a broad cross-correlation peak with a 50 day half 
width. The broad peak is punctuated by sub-peaks as already noticed by
Schild and Thomson, who comment that the cross-correlation sub-peaks 
``... tend
to have a uniform spacing of 16 days, which may correspond to an internal
reflection within the inner quasar structure.'' 

Everybody agrees that the central object must be illuminating the inner
edge of the accretion disc, and that since we are dealing with high
luminosity quasars one should use the Shakura-Sunyaev (1973) thin accretion
disc model. For the case of a central black hole in Q0957, this would be at
a radius less than or equal to $6 R_G$, which is less than a light day for
either quasar system, even with
cosmological (1+z) correction to the observer's clock. The fact that a
luminosity associated with the inner edge of the accretion disc is not
observed in cross-correlation, or autocorrelation, or microlensing for Q0957
means 
that there is nothing there. What is seen instead is the structure in the
autocorrelation plot with sub-peaks around the 417-day peak of the overall
autocorrelation curve, Fig. 5.

These sub-peaks are always around 10 days away from the cosmological
time delay, 417 days, and so they are easily understood as the bright inner
edge of the accretion disc that everybody agrees must exist. But this makes
the inner quasar structure approximately 10 times larger than expected. 
With elaborate corrections for the
inclination of the object, SLR06 found the luminous accretion disc inner edge
to have a radius $\sim 70 R_G$ in Q0957, instead of the $\sim 6 R_G$ 
expected for a black hole. 
The sub-peaks in the Fig. 5 autocorrelation plot are around 
379, 392, 408, 417, 420, and 425 days.

These cross-correlation experiments have long shown an interesting effect,
namely that the cosmological time delay measured for this data has been
changing with time. This was discovered in Thomson \& Schild (1997) where
contours of the cross-correlation time delay as a function of observation
date are shown. It may be seen that the time delay peak was shortening with
calendar date. 
This may be understood as a result of the
magnification cusp of a microlensing star passing across the face 
of this quasar, and
highlighting the internal structure progressively. 

Thus from many points of view we infer that if there is any
luminosity within $70 R_g$ we will have seen it selectively amplified at some
time. Instead, we just seem to see those $70 R_g$ structures being selectively
amplified at time scales near 10 days.
Moreover, hypothetical structure at $6R_G$ orbiting at the inner
edge of the accretion disc, would be expected to create strongly periodic 
brightness fluctuations, not observed. On the basis of these observations
it was concluded in SLR06 that an intrinsic magnetic propeller contained
within the central compact object of Q0957 had cleared out the
inner region of the accretion disc. This was found to be consistent with
the MECO model while being inconsistent with inner structure at $6 R_G$
inferred for the black hole model. The existence of internal structure is
probably part of the reason why seemingly reliable calculations do not
consistently indicate the same time delay even for data sets considered
reliable by Oscoz et al (2001).

The published SLR06 report contains an explanation for the cross-correlation
peaks at the particular values around 392, 404, 424 days found, for the 
already determined inclination of the quasar and inner structure at a
radius $\sim 70 R_G$ . 
The discussion
in Section 2 of SLR06, which we do not repeat here, shows that inner 
luminous ring structure produces
cross-correlation at a variety of lags as observed, since the nearest and
farthest surfaces would both produce strong reflection.

Also discussed in SLR06 is the location of the radio emitting region. As
demonstrated there, reverberation of the compact core radio emission has
been found to be 30 days (observer's clock, UV-optical leading radio) and a
series of microlensing events of the radio emission allow the size and
fraction of the compact radio source emission to be determined from the
duration and radio brightness amplitude of the events.

\section{The Implications of Rapid Quasar Microlensing}

The preceding sections have detailed why we believe that evidence exists for
quasar luminous continuum structure on size scales of the outer Elvis
structure, 50 proper light days, and for an inner structure $\sim 70 R_g$. 
But another observation shows an important characteristic of the inner
structure. 

In Fig. 6, which is a repeat of Colley and Schild (2003), we show a 
simple plot of the quasar
brightness measured unusually carefully for the duration of the night, in
1995.9 and again 417 days later. In this plot, brightness measured for the
first arriving A image is plotted as open circles, with a single data point
and error bar plotted for the hourly averaged brightness. It may be seen
that nightly brightness trends with typical amplitudes of .01 magnitude 
were found for the A image. 

The brightness of the B image at the same quasar time, measured at Earth
417.1 days later and plotted as filled circles, shows evidence for the 
same pattern of brightness trends. Thus on
JD-2449701.9 the A image underwent a 0.01 magnitude fading and the B
image recorded 417 days later shows that the A fading was a continuation of
a quasar fading trend that had begun several hours earlier. Similar trends
observed on the 3 following nights provide simple evidence that the quasar
has intrinsic brightness drifts with amplitude 0.02 magnitudes on time
scales of 12 hours, that are being recorded with reasonable accuracy and
seen at Earth with the 417.1 day cosmological time delay lag.
And that the 417.1 day time delay found from a continuous monitoring
campaign reported in Colley et al (2003) must be correct for this 
data sample.

Now we remark on the events recorded on JD - 2449706, where image A had a
deep (2\%) minimum and image B, observed 417 days later, had a shallow
minimum. The difference between the two brightness trends must have been
caused by microlensing. A more convincing plot of this observed event is
shown as Fig. 2 in Colley and Schild (2003).
This simple observation of a 1\% microlensing brightness change in 5 hours
proper time challenges theory on two
points; it is not understood in the black hole model how the quasar can
change brightness due to intrinsic quasar fluctuation processes for
quoted accretion disc sizes, or how microlensing can change so quickly for
the quoted quasar luminous accretion disc sizes. And for microlensing to
occur on such short time scales, a fine structure in the microlensing
caustic pattern due to a graininess of the mass distribution in the lens
galaxy from planet mass microlenses must be present. Comparable fluctuations
have been simulated by Schild and Valukik (2003) with equal positive and
negative sub-cusps on day time scales only if planet-mass microlenses
dominate the mass of the lens galaxy.

We emphasize that the rapid microlensing event requires two physical
effects not expected in astronomy; the quasar must have fine structure, and
the mass distribution within the lens galaxy must be dominated by planetary
mass bodies. Direct simulations of the Q2237 microlensing by Vakulik et al
(2007) also conclude that the microlensing must be caused by Jupiter mass,
$10^{-3}M_{\odot}$, compact objects.

Standard arguments of causality require that the existence of observed
fluctuations intrinsic to the quasar imply structure on scales of 12 light
hours, which after correction for cosmological (1+z) time dilation
corresponds to a size scale of less than $1 R_g$. With the spherical or
cylindrical geometry assumed for the quasar central source, this implies
the existence of a ring structure dimension of such size, and we have
interpreted this as the radial thickness of the accretion disc inner edge,
since that would be the smallest dimension associated with an accretion
disc model. Thus the rapid fluctuations seen convincingly in Fig. 6 and
found previously in the data and structure function, Figs. 3 and 5, of 
Colley \& Schild (2000)
probably imply the existence of quasar structure with significant
luminosity on a small spatial scale of $1 R_G$.

Further analysis in section 2 of SLR06 amplifies this result, 
with the additional conclusion that the
linear increase in the wavelet amplitude with lag found by Schild (1999)
is also compatible with a ring structure for the central emitting region.

\section{Summary and Model Comparison}

The purpose of this section is to summarize the results from each
preceding section, and to comment upon the relevance to the fundamental
issue, which is whether these first results of the direct detection of
UV-optical and radio inner quasar structure favor a black hole or MECO
interpretation. 

In section 2, we considered the quantity and quality of data that
contribute to the conclusion that Q0957 does not have accretion disc
structure expected at approximately $6R_G$ (somewhat smaller for rapid
rotation of the central object). We showed that hundreds of data points
define the autocorrelation  and cross-correlation estimates of
structured quasar brightness trends, and their 1 \% accuracy was
emphasized. 

In section 3 we emphasized that autocorrelation estimates for long lags, up
to 2 years, showed structure that suggested that any disturbance seen first
in the central region reverberates in outer structure that has size scales
attributed previously to the Elvis outflow structures. Originally
discovered in Thomson and Schild (1997) these structures now are
interpreted to reveal the details of the quasar's central structure and
orientation in space (Schild 2005). 

We also found in section 3 that autocorrelation plots revealed the presence of
UV-optical luminous inner structure not at the location of the innermost
stable orbit (less than or on the order of $6R_G$) but rather at a much 
larger radius on the order of
$70R_G$. This observation, consistent with the intrinsic magnetic propeller
model associated with the MECO theory for quasar structure (SLR06), is not
that which is expected from black hole quasar structure models. 
Figures 3 and 4 seem to indicate the existence of inner structure,
and the width of the central autocorrelation peak (25 days, observed) 
is interpreted
as a determination of the structure's central radius in light days.

In section 4 we recalled that many cross-correlation calculations have sought
a sharp peak for the cosmological time delay, but it has not been found for
any of the long brightness time-series records available. This is most
easily understood as the result of the internal structure and its
significant microlensing, and evidence for this interpretation is probably
seen in the fine structure of the cross-correlation estimates. The evident
cross-correlation fine structure is arguably on the time scale of the inner
structure evident in the auto-correlation plots, Figs 3 and 4, with
evident structure at lags 10 - 20 days.

In section 5 we examined the evidence for a rapid microlensing that is seen
to 99.9\% significance in data obtained in two separate observing seasons.
The generally good agreement of the data from the two seasons within the
quoted error bars for overlapping data points provides confirmation that
the brightness estimates and adopted time delay
are correct. And the existence of such rapid
microlensing requires the existence of sharp quasar structure that is
incompatible with black hole physics and was not predicted before it was
observationally discovered.

With these key observational results in mind, we discuss the comparison of
standard black hole models to MECO model results. Although a general
relativistic 3-dimensional simulation of a MECO object has not yet been
undertaken, an analytical model has been created by Robertson and Leiter
(2002, 2003, 2004, 2005) and is the basis for these conclusions.

Based on the analysis of the observational evidence presented in this paper
it has been shown in SLR06 that Q0957+561 has the four intrinsic structural 
elements as follows:
\begin{itemize}
\item 1. Elliptical Elvis Structure located at distance Re from the central
  object and height He above the accretion disc plane:  $Re = 2 \times 
  10^{17}$cm (320 $R_G,$ 77 light days), He = $5 \times 10^{16}$cm (80 $R_G,$
  19 light days).
   
\item 2. Inner Radius of Accretion Disk: $R_{disk} = 4 \times 10^{16}$cm =
  $64 R_G$ = 15 light days.
   
\item 3. Hot Inner Accretion Disk Annulus:  delta(R) = $5.4 \times
  10^{14}$cm = 1 $R_G$ = 5 light hours. 
   
\item 4. Base of Radio Jet: $R_{rad} = 2 \times 10^{16}$cm (8 light days), 
  $H_{rad} = 9 \times 10^{16}$cm (35 light days). 

\end{itemize}
A cartoon illustrating a cross-section of this quasar structure model has
been given as Fig. 1 in SLR06. Attempts to explain all four components of 
the observed
inner structure of the quasar Q0957 in terms of standard central black hole
models have failed for the following reasons:

a) Modeling them in terms of an intrinsic magnetic moment generated by a
central spinning charged black hole fails because the necessary charge on
the spinning black hole required would not be stable enough to account for
the long lifetime of the inner quasar structure.

b) Modeling them in terms of a Kerr black hole ADAF accretion-
disk-corona­jet (Narayan \& Quataert 2005; McKinney \& Narayan 2007)
in which the magnetic field is
intrinsic to the accretion disk and not intrinsic to the central rotating 
black hole,
fails because it cannot account for the very large opening angles for the
coronal Elvis outflows, and in particular it cannot account for the hot thin
inner disk annulus that is observed

c) Modeling them in terms of a Magnetically Arrested Disk (MAD) black hole
(Igumenschev et al. 2003) fails since it cannot account for the hot thin
inner disk annulus that is observed within the inner structure of Q0957.
Instead the MAD model predicts the existence of orbiting infalling 
hot blobs of plasma
inside the inner edge of the accretion disk that would produce periodic
brightness fluctuations, which are not observed.

However it has been shown that all the four
components of the observed inner structure in the quasar Q0957+561 can be
consistently described within the context of the Magnetospheric
Eternally Collapsing Object (MECO) model described in SLR06, 
in which a very strong intrinsic magnetic field anchored to
a highly redshifted rotating central compact MECO interacts in a magnetic
propeller mode with the surrounding accretion disk and generates all the
four components of the Q0957 structure. Such MECO models are characterized by
highly redshifted, Eddington-limited, collapsing central compact objects
containing strong intrinsic magnetic fields aligned with their MECO axis of
rotation. 

The MECO contains a central rotating magnetic object whose dynamo sweeps
clean the central region of the quasar out to a distance at which the
magnetic propeller acts on the inner edge of the accretion disk, and a
radio-emitting region above the disk where magnetic field lines must twist
and bunch up until they eventually break and reconnect at relativistic
speeds. Such an object does not have an event horizon; instead, infalling
material collects at an inner structure just beyond 2Rg that further
collapses to higher redshift while remaining in causal connection for all
time. Because of the small light cone angle for radiation escaping from this
highly redshifted region to the distant observer, the resulting low
luminosity in the far-infrared wavelengths makes this region difficult to
detect.

In the MECO model for Q0957, the magnetic propeller interacts
with the inner regions of the accretion disk and creates a very thin hot
inner annular (band-like) structure and an outer coronal structure
characterized by strong relativistic outflow with a wide opening angle to
the z-axis of rotation as is observed. In addition the size and location of
the radio-emitting region associated with the structure in the quasar
Q0957+561 have been found to correspond to the region above the central
compact object where the reconnection of magnetic field lines at
relativistic Alfven speeds, like that generated by a rotating central MECO
containing an intrinsic magnetic field, should occur. The structures found
to dominate the UV-optical and radio emission are shown correctly scaled in
Figure 7.

It is important to note that the MECO model that best fits the inner
structure observed in the quasar Q0957+561 differs significantly from most
black hole models currently under consideration. In particular, the observed
structure they generate seems to resemble the complex inflow-outflow pattern
seen in magnetic propeller models for young stellar objects. The action of
such magnetic propeller forces on young stellar objects has been discussed
and simulated by Romanova et al. (2002, 2003a, 2003b) with non-relativistic
models that produce observable structures whose spatial geometry is very 
similar to the inferred Schild-Vakulik (2003) structure.

On the basis of the above arguments we come to the conclusion that
observation of the four components of the stable non-standard inner
structure within the quasar Q0957+561, and especially the existence of the
hot thin inner disk annulus that is observed, represents strong evidence for
the existence of an observable intrinsic magnetic moment, generated by a
supermassive $3-4 \times 10^9  M_{\odot}$ MECO acting as the 
central compact object in this
active galaxy, which implies that this quasar does not have an event
horizon.

\section{Concluding Remarks}

We have shown that a comprehensive analysis of the gravitational lensing
and microlensing observations of the quasar Q0957 indicates that the
internal structure of this quasar appears to be dominated by a highly
redshifted supermassive central compact object that contains an observable
intrinsic magnetic moment and therefore cannot be described by a black
hole. The implications of this startling discovery are so profound that it
is necessary to clarify and review the supporting observational evidence.

For the benefit of the astrophysics community,
this paper has summarized within a single document the full list of unique 
gravitational microlensing and reverberation observations of
quasar Q0957 which have permitted a detailed reconstruction of the intrinsic
structures emitting radiation from its interior regions. Surprisingly these
observations failed to reveal the expected accretion disk extending in close
to the central object. Instead it was found that the inner accretion disk
contained a large empty region which ended at a large radius where we found
a thin hot inner ring. In addition, it was found that there was a large
hyperbolic Elvis outflow structure about ten times further out.

The four components associated with these internal structures were found to
be similar to the features revealed in simulated accretion flows into the
central magnetic dipole objects contained within Young Stellar Objects,
which have been successfully simulated by Romanova et al.
Hence our main conclusion was that we were seeing a similar type of central
magnetic object in this quasar and if so, such an object cannot be a black
hole because black holes cannot possess an intrinsic anchored magnetic
dipole field. Hence these observations were found to represent strong
evidence for the existence of a new kind of central collapsed object in the
quasar Q0957 called a "Magnetospheric Eternally Collapsing Object" (MECO)
which is permitted within the framework of general relativistic gravity
(SLR06).

MECO form by the same gravitational collapse process believed to result in
black holes, but due to internal Eddington limited radiation pressure, they
are never observed to collapse through an event horizon (Mitra, 2006). 
Due to the extreme
surface redshift of a MECO, its surface radiation is too faint to be easily
detected at astronomical distances. Hence MECO differ observationally from
black holes primarily by the ability of the MECO to exhibit observable
manifestations of its intrinsic magnetic field on its surrounding accretion
disk and environment.

While many models of quasars based on black holes exist in the literature,
none of them are able to account all four of the components of  the internal
structure observed within Q0957. This was especially true for the observed
inner accretion disk structure, which contained a large empty region that
was truncated at a large radius by a very thin hot inner ring (illustrated
white in Figure 7) characteristic
of an intrinsic magnetic propeller interaction with the accretion disc. 
Hence we conclude in response to the question asked in the title of this 
report, that the observational evidence for the existence of a very hot thin
luminous ring deep within quasar Q0957 represents a strong observational
argument in favor of the existence of a supermassive MECO in the center 
of this quasar, instead of a black hole.

\section{Acknowledgements}

This report is the result of colloquium and extended discussions with Alan
Bridle at the National Radio Astronomy Observatory (NRAO) in
Charlottesville, Virginia. Both authors  thank him for encouraging 
the development of this paper by offering constructive criticism and
suggesting important reading topics. In particular, Darryl Leiter thanks
him for allowing full access to NRAO visiting scientist facilities during
the time period when this report was written.
Most of the statistical calculations reported here were
undertaken by David J. Thomson on the CRAY YMP computer, formerly at AT\&T
Bell Labs. Rudy Schild thanks Professor Thomson for countless discussions 
about the subtleties of the Q0957 brightness record.

\newpage
\begin{figure}
\begin{center}
\plotone{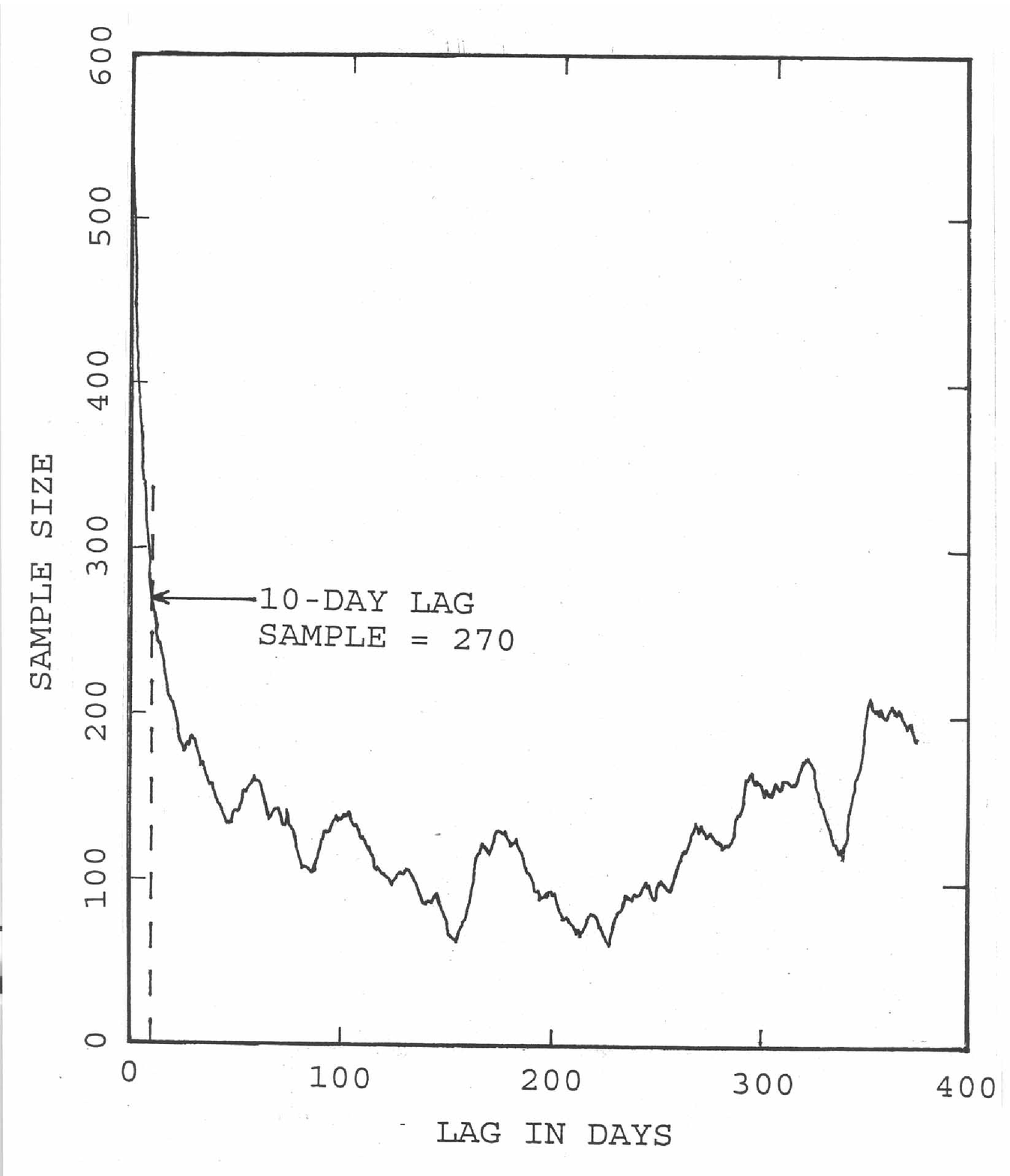}
\caption{The data sample size emphasizing the number of nights contributing
to correlation estimates of quasar brightness. As described in the text,
interpolations of up to 2 days were included to describe the quantity of
data contributing to correlation calculations. For the present example,
data measured first in image A and compared to the later arriving B image,
measure in the hundreds, for lags relevant to quasar structure, as well as
to microlensing.}
\label{fig. 1}
\end{center}
\end{figure}

\newpage
\begin{figure}
\begin{center}
\plotone{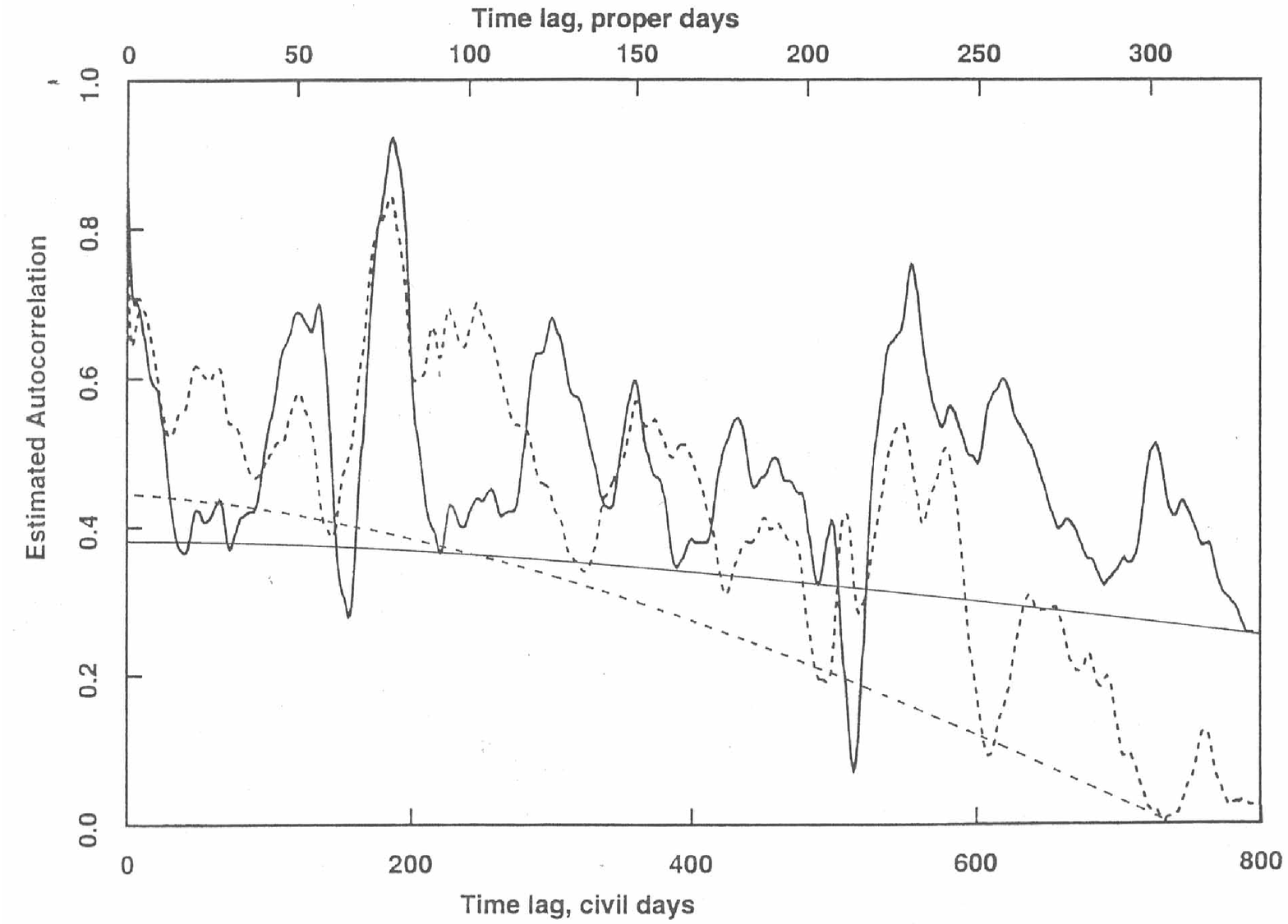}
\caption{An autocorrelation plot for the A (dashed curve) and B(solid curve)
quasar images. While previously this data
analysis was used to infer the quasar outer structure, we now
emphasize the evidence for inner structure, for lags less than 20 days
(civil days). The central peak, near 0 lag, does not have the rapid falloff
with a time scale near the sampling time expected for noise-dominated
data. Instead it shows structure on time scales near 20 days, understood as
indicative of inner quasar structure.}
\label{fig. 2}
\end{center}
\end{figure}

\newpage
\begin{figure}
\begin{center}
\plotone{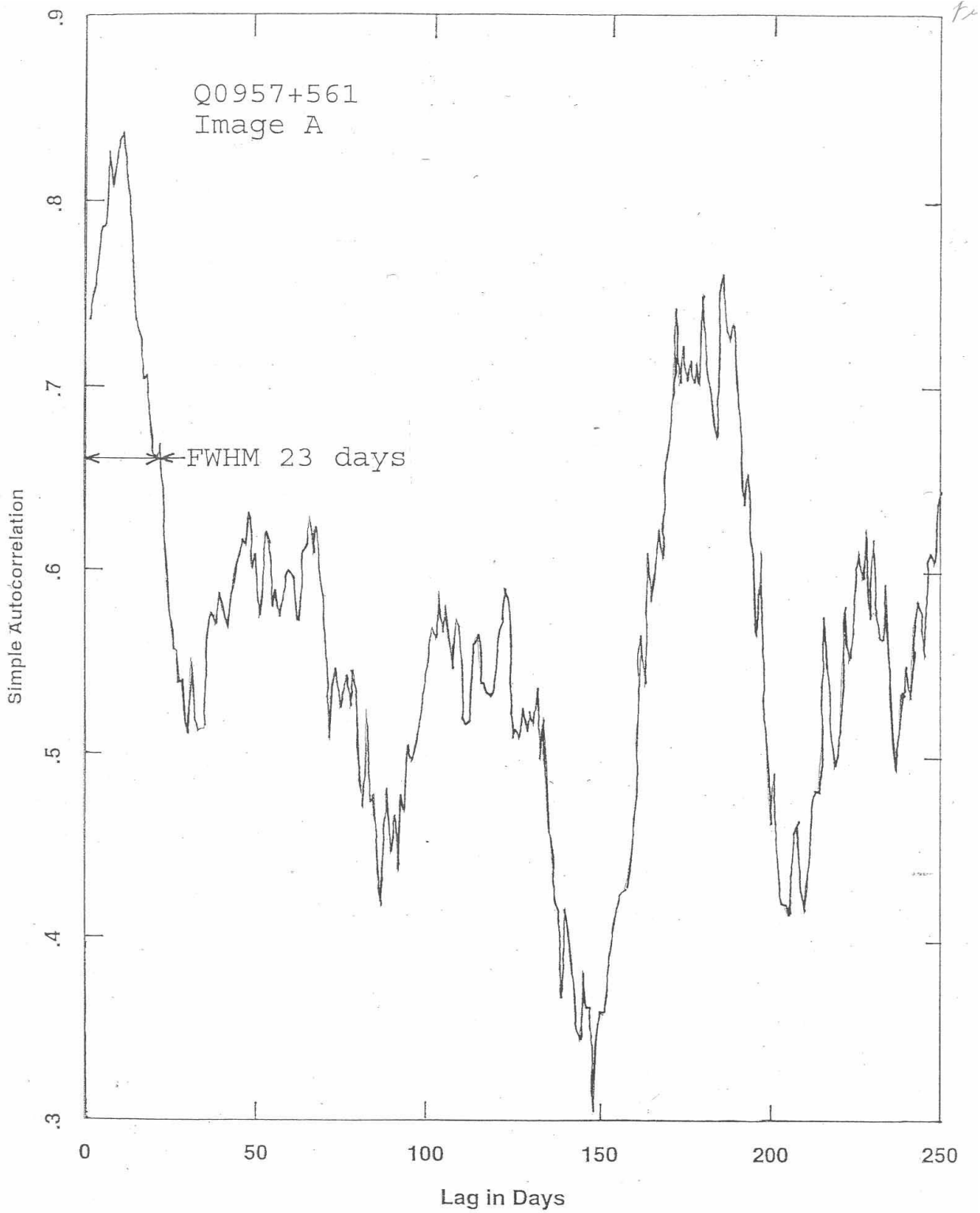}
\caption{An autocorrelation plot of the image A data with finer resolution
for smaller lags. Prominently seen are the lags for outer Elvis structures,
as well as structure from the central peak with an overall half-width of 23
days and finer sub-structure at 6, 11, and 20 days.}
\label{fig. 3}
\end{center}
\end{figure}

\newpage
\begin{figure}
\begin{center}
\plotone{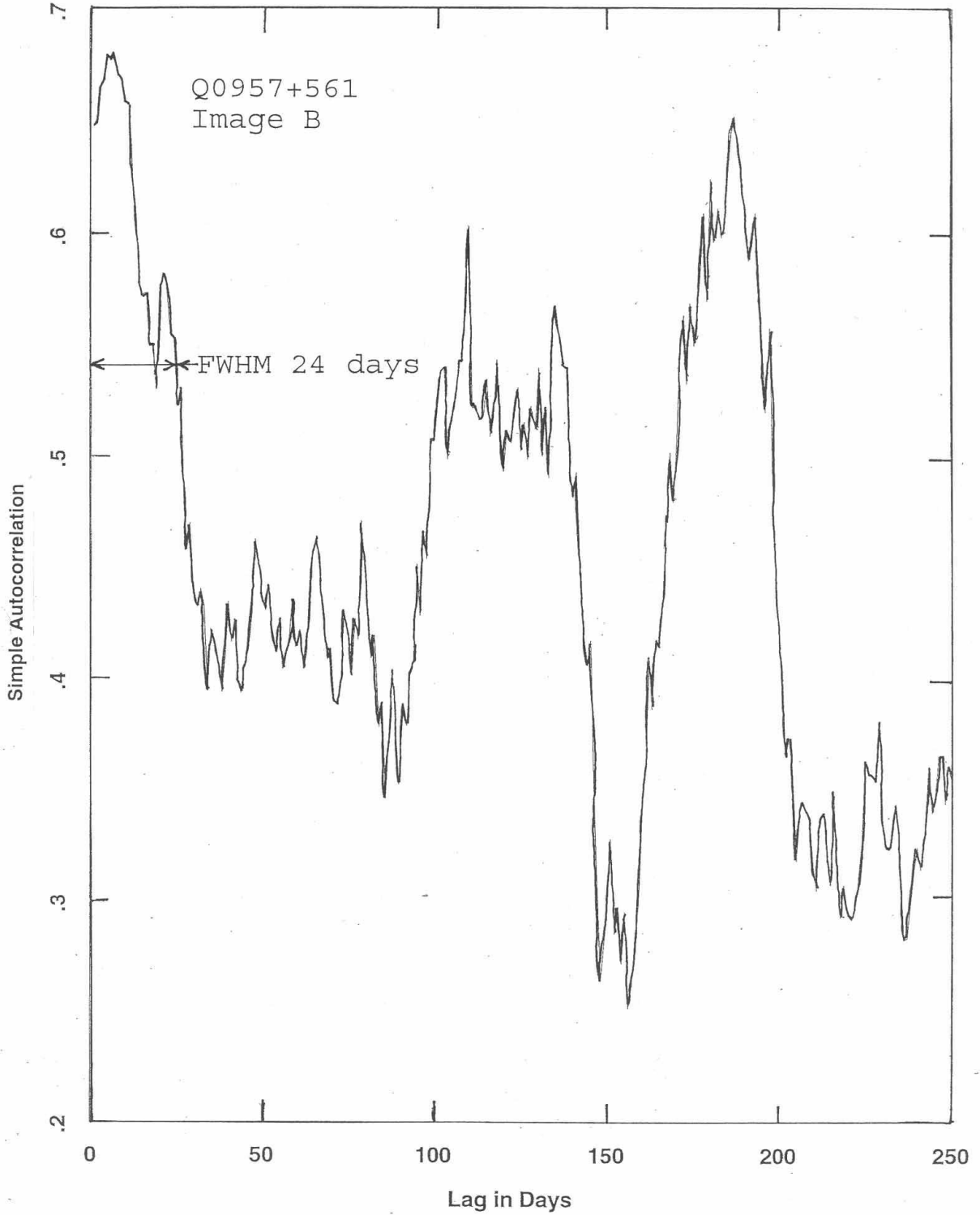}
\caption{A fine structure autocorrelation plot for the B quasar image.
Strong autocorrelation peaks at 129 and 192 days from the outer Elvis
structures are evident. Also well seen is the structure surrounding the
central peak and evidencing central quasar structure. The overall width
is approximately 24 days, and substructure peaks are found at 5 and 20
days.} 
\label{fig. 4}
\end{center}
\end{figure}

\newpage
\begin{figure}
\begin{center}
\plotone{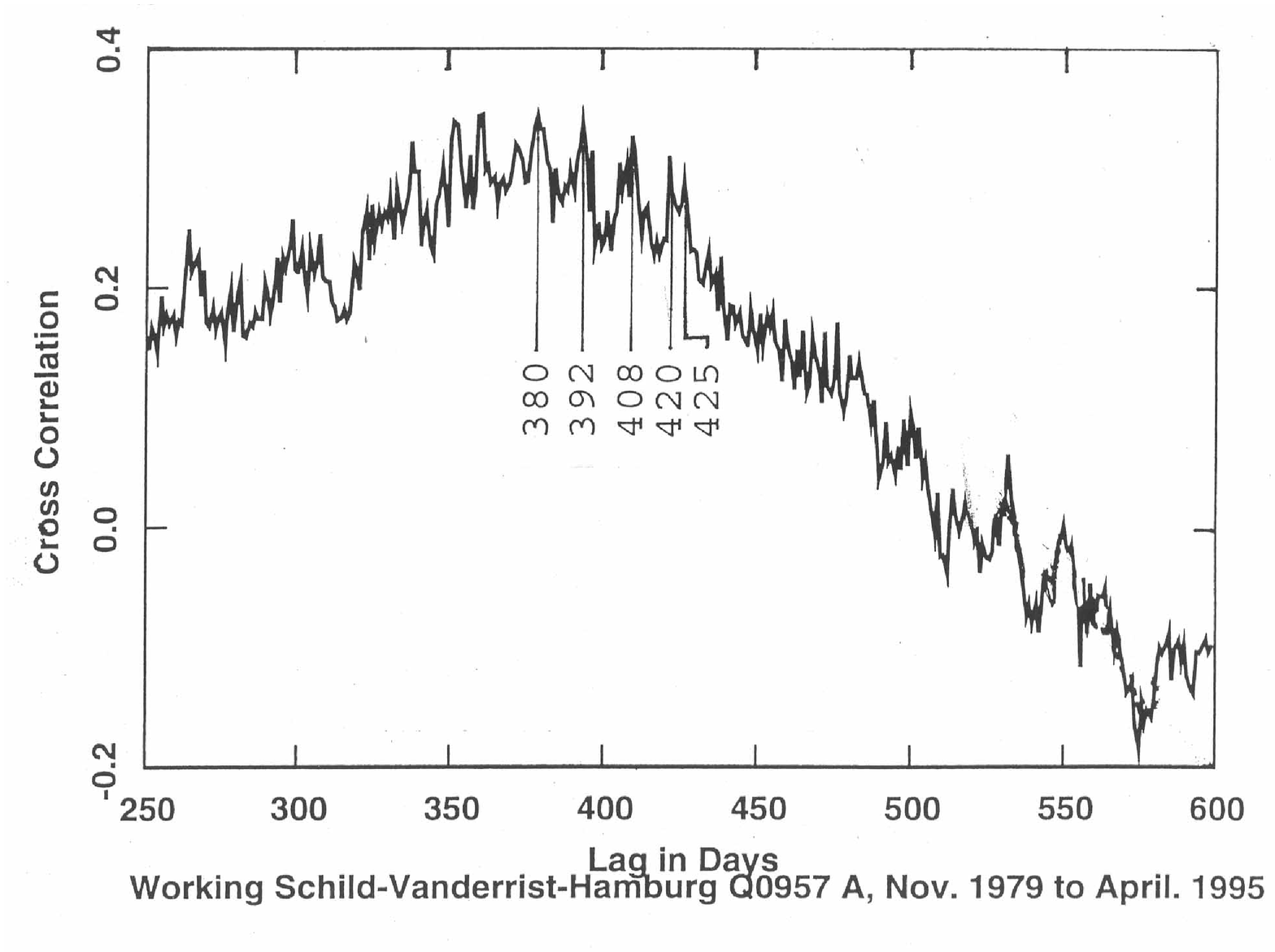}
\caption{A cross-correlation plot centered on the value of the cosmological
time delay, but showing the complications from quasar structure. In a
simplistic view with no influence of quasar structure, the
cross-correlation should be a strong peak at the cosmological value, 417
days. Because of quasar structure, a broad peak with FWHM near 100 days is
found instead. Fine structure within this broad peak is evidently
indicative of the quasar's inner structure.}
\label{fig. 5}
\end{center}
\end{figure}

\newpage
\begin{figure}
\begin{center}
\plotone{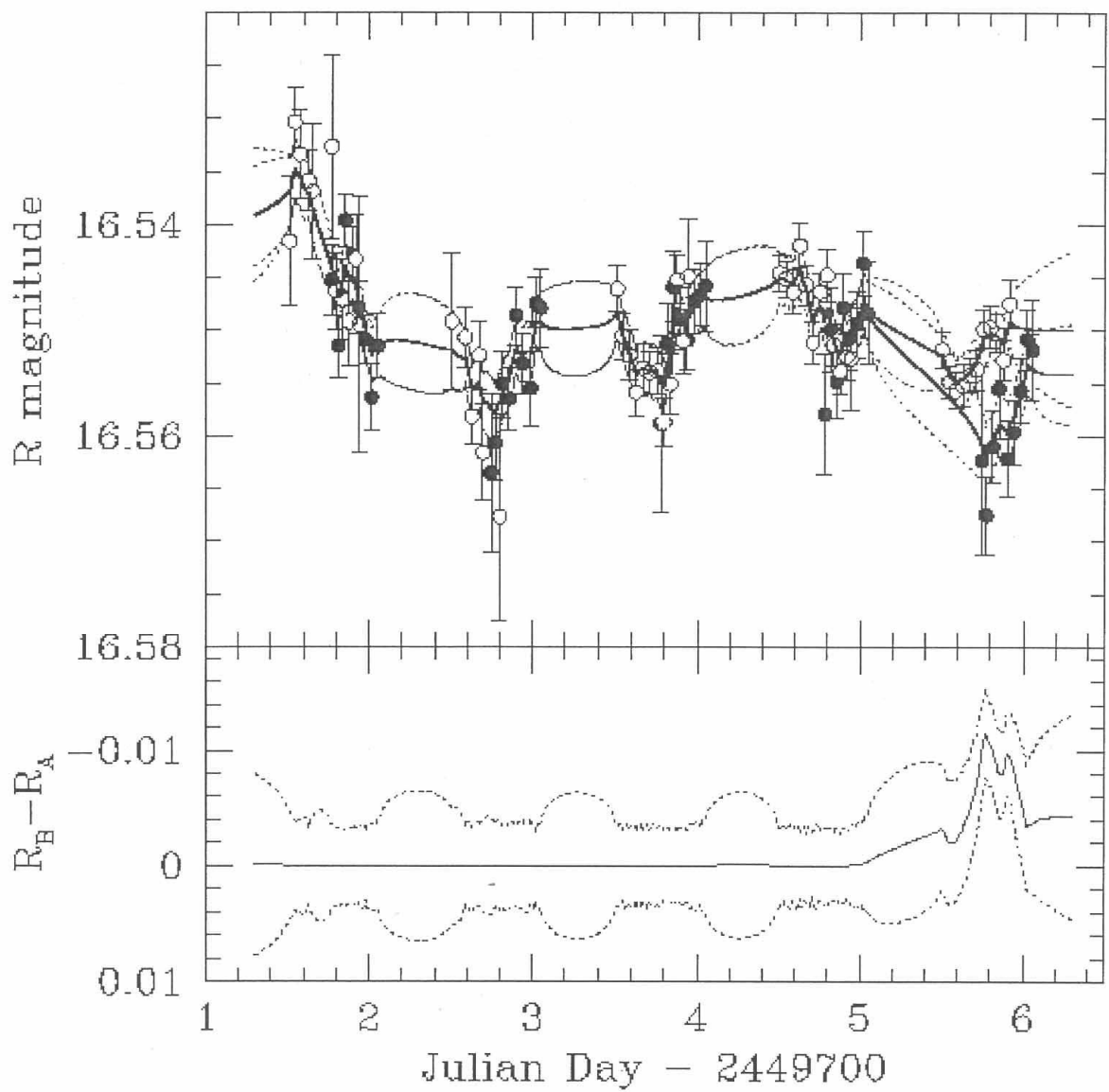}
\caption{A simple plot of R magnitude measured for image A (open circles)
in 1995.9 and in 1997.1 for image B (filled circles). A pattern of
well-defined quasar brightness variations seems indicated for the first
four nights. But on the fifth night, the brightness records significantly
diverge, meaning that microlensing by a planet mass compact object in the
microlensing galaxy made image B fainter on a time scale of hours. In the
lower panel, the smoothed time-delay-corrected A-B magnitude is plotted
with linear interpolation, and with 1 $\sigma$ outer limits per hourly 
data point shown as a dotted line. We see that the rapid microlensing event
at JD 2449706 was securely observed as probably an event where the quasar
faded and the microlensing also diminished.}
\label{fig. 6}
\end{center}
\end{figure}

\newpage
\begin{figure}
\begin{center}
\plotone{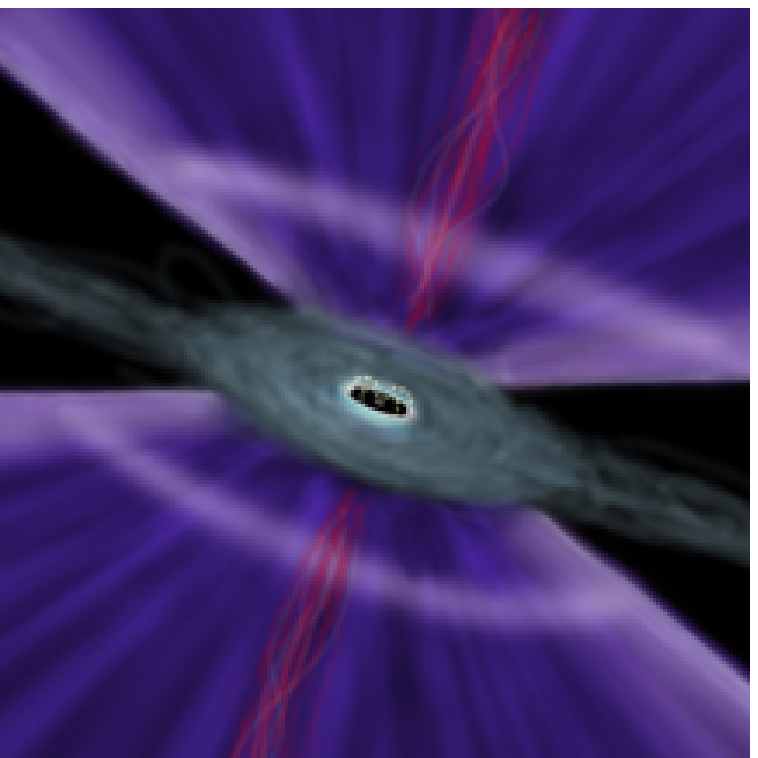}
\caption{ A schematic figure demonstrating the principal luminous quasar
structures as determined by our reverberation-microlensing analysis. 
The dark compact central object is surrounded by dipole field lines (dotted
yellow) and the sharp luminous ring at the inner edge of the accretion disc
is white. A dark accretion disc intersects the outflow wind structures
(Elvis surfaces) whose fluoresence above and below the plane (blue)
contributes 
to the UV-optical continuum observed. The compact radio core (red) is 
shown in size and distance scaled to the overall structure. }
\label{fig. 7}
\end{center}
\end{figure}


\newpage

\begin{thebibliography}{}
\bibitem{CS00} Colley, W. \& Schild, R., 2000, ApJ, 540, 104
\bibitem{CS03} Colley, W. \& Schild, R., 2003, ApJ, 594, 97 (CS03)
\bibitem{CS+} Colley, W. et al, 2003, ApJ, 587, 71
\bibitem{INA} Igumenschev, I. Narayan, R. and Abramowicz, M.,
2003, ApJ, 592, 1042
\bibitem{McN} McKinney, J. \& Narayan, R. 2007, MNRAS, 375, 513
\bibitem{am} Mitra, A., 2006, New Astronomy, 12, 146
\bibitem{NQ05} Narayan, R. \& Quataert, E. 2005, Science, 307, 77 
\bibitem{Osc01} Oscoz, A. et al, 2001, ApJ, 552, 81
\bibitem{R64} Refsdal, S. 1966, M.N.R.A.S., 128, 307
\bibitem{RL02} Robertson, S., and Leiter, D. 2002, ApJ, 565, 447
\bibitem{RL03} Robertson, S., and Leiter, D. 2003, ApJ, 596, L203
\bibitem{RL04} Robertson, S., and Leiter, D. 2004, MNRAS, 350, 1391
\bibitem{RL05} Robertson, S., and Leiter, D. 2005, `` The Magnetospheric
Eternally Collapsing (MECO) Model of Galactic Black Hole Candidates and
Active galactic Nuclei,'' pp1-45 (in New Directions in Black Hole
Research, ed. P.V.Kreitler, Nova Science Publishers, Inc, ISBN
1-59454-460-3, novapublishers.com
\bibitem{R02}  Romanova, M. et al, 2002, ApJ, 578, 420
\bibitem{R03a}  Romanova, M. et al, 2003a, ApJ, 588, 400
\bibitem{R03b}  Romanova, M. et al, 2003b, ApJ, 595, 1009
\bibitem{R04}  Romanova, M. et al, 2004, ApJ, 616, L151
\bibitem{S91} Schild, R., 1990, AJ, 100, 1991
\bibitem{S90} Schild, R. E. 1996, ApJ, 464, 125
\bibitem{S99} Schild, R. E. 1999, ApJ, 514, 598
\bibitem{S04b} Schild, R. E. 2004a, astro-ph/0409549
\bibitem{S04a} Schild, R. E. 2004b, astro-ph/0406491
\bibitem{S05} Schild, R. E. 2005, AJ, 129, 1225
\bibitem{S07} Schild, R. E. 2007, astro-ph/07082917
\bibitem{SC86} Schild, R. E., \& Cholfin, B. 1986, ApJ, 300, 209
\bibitem{ST97} Schild, R. E., \& Thomson, D.J. 1995, AJ, 109, 1970
\bibitem{ST97} Schild, R. E. \&  Thomson, D.J. 1997, AJ, 113, 130
\bibitem{ST97a} Schild, R. E., \& Thomson, D. J., 1997a, AJ, 113, 130
\bibitem{SV03} Schild, R. E. \& Vakulik,V. 2003, AJ, 126, 689
\bibitem{SD07} Schi;d, R. E. \& Dekker, M. 2007, A.N. 327, 729
\bibitem{SLR06}Schild, R. Leiter, D. \& Robertson, S. 2006, AJ, 132, 420
(SLR06) 
\bibitem{slr07}Schild, R. Leiter, D. \& Robertson, S. 2007,
astro-ph/07082422 
\bibitem{ss73} Shakura, N. \& Sunyaev, R. 1973, A\&A, 24, 337
\bibitem{TS97} Thomson, D. J., \& Schild, R., 1997,, in Applications of
         Time Series Analysis in Astronomy
and Meteorology, ed. T. Subba Rao, M. Priestly, \& O. Lessi
[Chapman and Hall: New York], 187
\bibitem{VA06}Vakulik, V. et al, 2006, A\&A, 447, 905
\bibitem{VA07}Vakulok, V. et al, 2007, astro-ph/07081082
\bibitem{wcs79}Walsh, D. Carswell, R. and Weymann, R. 1979, Nature, 279, 381
\bibitem{Woz00} Wozniak, H. et al, 2000, ApJ 529, 88

\end{thebibliography}
\end{document}